\documentclass[%
reprint,
superscriptaddress,
amsmath,amssymb,
aps,
]{revtex4-1}

\usepackage{graphicx}
\usepackage{float}
\usepackage[caption=false]{subfig}
\usepackage{dcolumn}
\usepackage{bm}

\begin{document}
\preprint{APS/123-QED}
\title{Impact of a non-uniform charge distribution on virus assembly}

\author{Siyu Li}
\author{Gonca Erdemci-Tandogan}
\author{Jef Wagner}
\affiliation{Department of Physics and Astronomy,
University of California, Riverside, California 92521, USA}
\author{Paul van der Schoot}
\affiliation{Group Theory of Polymers and Soft Matter, Eindhoven University of Technology, P.O. Box 513, 5600 MB Eindhoven,
The Netherlands}
 \affiliation{ Institute for Theoretical Physics,
   Utrecht University,
   Leuvenlaan 4, 3584 CE Utrecht, The Netherlands}
\author{Roya Zandi}
\affiliation{Department of Physics and Astronomy,
   University of California, Riverside, California 92521, USA}





\date{\today}

\begin{abstract}
Many spherical viruses encapsulate their genome in protein shells with icosahedral symmetry. This process is spontaneous and driven by electrostatic interactions between positive domains on the virus coat proteins and the negative genome. We model the effect of the icosahedral charge distribution from the protein shell instead of uniform using a mean-field theory. We find that the non-uniform charge distribution strongly affects the optimal genome length, and that it can explain the experimentally observed phenomenon of overcharging of virus and virus-like particles.
\end{abstract}

\maketitle



The simplest viruses consist of two components: the genome, either an RNA or DNA polynucleotide that carries the genetic code; and the capsid, a protein shell that encloses the genome. The capsid consists of many identical (or nearly identical) copies of the coat protein subunit. Even though the coat proteins are highly irregular in shape, the protein shells of most spherical viruses are highly structured and form icosahedral symmetry \cite{Bancroft,bruinsma,Fejer:10,Rapaport:04a}. One of the consequences of icosahedral symmetry is that it puts restrictions on the number of proteins that can make up a spherical virus shell. It limits this number to 60 times the structural index $T$ that almost always assumes certain ``magic'' integer values $T=1,3,4,7,\ldots$ \cite{Wagner2015956,Chen:2007b,Stefan}.

Many small single-stranded RNA or ssRNA viruses have been shown to spontaneously self-assemble {\it in vitro}, that is, outside living cells in solutions containing virus coat protein subunits and genome. In fact, virus coat proteins are able to co-assemble with a variety of cargos such as RNAs of other sometimes unrelated viruses, with synthetic polyanions, and with negatively charged nanoparticles \cite{Sun2007,Kusters2015,Zandi2016}. The assembly of properly structured viral capsids in many icosahedral RNA viruses with the variety of cargoes is due to the presence of a disordered RNA binding domain on the N- or C-terminal end of the protein subunits rich in basic amino acids that extend into the capsid interior. These basic amino acids are positively charged under most solution conditions and adorn them with either a few to tens of positive charges depending on the species of virus. It is now widely accepted that electrostatic interactions between the positive charges on the coat protein tails and negative charges on the genome is the driving force for the spontaneous assembly of simple viruses in solution \cite{Cornelissen2007,Ren2006,Bogdan,Anze2,DanielDragnea2010,Zlotnick,Hsiang-Ku,Venky2016}.

Na\"{\i}vely, one might expect that the total charge on the genome and the capsid should balance out, if not perfectly then certainly approximately. However, in many ssRNA viruses the number of negative charges on the genome significantly exceeds the number of positive charges on capsid proteins. For example, the number of positive charges on capsids of Cowpea Chlorotic Mottle Virus (CCMV) and Brome Mosaic virus (BMV), both with $T=3$ structures, is about 1800, yet their genome measures about 3000 nucleotides (\textit{nt}) \cite{Comas}. As each nucleotide bears a single charge, this suggests an \textit{overcharging} of over 60 per cent. Furthermore, in a recent set of {\it in vitro} experiments, where shorter segments of BMV RNAs in the range of  $500 - 2500$ \textit{nt}s were mixed with CCMV capsid proteins, the resulting virus-like particles (VLPs) had a mixed population of pseudo $T = 2$ and $T = 3$ shells that were all overcharged \cite{Comas}. RNA molecules shorter than 2000 \textit{nt}s were packaged in multiple copies, $e.g.$, four in the case of $500$ \textit{nt} RNAs or two for $1000$ \textit{nt} RNAs in pseudo $T=2$ capsids and two $1500$ \textit{nt} RNAs in $T = 3$ capsids.

While the {\it in vitro} self-assembly studies show that RNA-based virus-like particles are overcharged, experiments with linear negatively charged polymers rather than virus RNAs are less conclusive. In fact, studies with linear polyanions, such as poly(styrene sulfonate) (PSS), have often focused attention more strongly how the capsid size distribution is impacted upon by either the polymer length or the stoichiometry ratio of the capsid proteins and polymers \cite{Cornelissen2007,Hu2008}. What is known, is that polymers ranging in degree of polymerisation from $1900-16500$ monomers could all be encapsidated by a $T=3$ structure, resulting in anything from a weakly to a highly overcharged structure \cite{Hu2008}.  \textit{In vitro} self-assembly studies on mixtures of CCMV coat proteins and PSS chains as short as 180 monomers show a bimodal distribution of particle sizes corresponding to $T=1$ and pseudo $T=2$ structures\cite{Cadena2011}. According to these experiments there are on average two polymers in each $T=1$ capsid (600 positive charges) and three in each T=2 (1200 positive charges) \cite{Cadena2011}. Hence, the VLPs are in this case undercharged: the ratio of negative to positive charges is 0.6 for the $T=1$ and 0.45 for $T=2$ capsids. From these findings it is not easy to extract a sound conclusion about the optimal length of the encapsidated polymer. 

Several theoretical studies have shed light on the puzzling phenomenon of overcharging. Simulations on encapsulation of polymers with a fixed number of branching (quenched) as a model for RNA have shown that the level of overcharging is a sensitive function of the secondary and tertiary structures of the RNA \cite{elife, Ben-Shaul2015}.  Field-theoretical calculations presuming the branching to be annealed, not quenched, have also shown that the length of encapsidated polymer, and hence the level of overcharging, increases as the number of branch points increases \cite{Gonca2014,Gonca2016}.  While these theoretical studies confirm that the topology of RNA is important to the phenomenon of overcharging, they also predict that the optimal number of charges on a linear polyelectrolyte must be \textit{less} than the total number of charges on the inner capsid wall: these complexes must be under- rather than overcharged. This contrasts with the molecular dynamics simulations of Perlmutter et al., which show that even linear polyanion encapsulation can lead to overcharging \cite{elife}.

In virtually all theoretical studies focusing on the assembly of viral shells, the capsid has been assumed to be smooth and to have a uniform charge distribution in a region near the surface of the capsid \cite{bruinsma,Vanderschoot2009,Anze,Vanderschoot,Paul13a}. However, as already alluded to, in most simple RNA viruses the positive charges reside on the RNA binding domains of the coat proteins, which are arranged according to the underlying icosahedral symmetry of the shell. This implies that the charge distribution must somehow reflect this icosahedral symmetry, certainly near the surface of the capsid, and perhaps less so away from it. The effects of localization of charge near the inner surface of the capsid on the encapsulation of genome remain largely unexplored theoretically.

To remedy this, we study the impact of a non-uniform charge density on the optimal length of genome encapsulated by small icosahedral viruses.  Since $T=1$ and $T=3$ capsids have 60 and 180 RNA binding domains, respectively, we model capsids with 60 and 180 positively charged regions, as shown in Fig.~1. We show a non-uniform charge distribution, associated with the underlying icosahedral arrangement of the proteins part of a virus shell, results in a longer optimal genome length compared to a uniform charge distribution, and can give rise to the phenomenon of overcharging even for linear polyanions. The effects of a non-uniform charge distribution and the highly branched RNA secondary structures, typical for viral RNAs, combine to greatly enhance overcharging. This allows a larger amount of RNA to be packed in the same restricted interior of the virus shell, which would be an evolutionary advantage to the virus.

Furthermore, we find that the optimal length of the genome, and as a result that of the number of encapsulated charges, depends on the detailed structure of RNA binding domains, {\it i.e.}, the thickness, height and charge density. This is consistent with the experimental findings of Ni \textit{et al.} on Brome Mosaic virus (BMV), in which mutations in the RNA binding domains that keep the number of charges constant but change their length and charge density impact upon the packaged RNA length \cite{Bogdan}. Our theoretical calculations allow us to single out the impact of length and charge density of the RNA binding domains, without considering other effects such as the impact of translational entropy and kinetic trapping that make the interpretation of experiments and simulations difficult.

Our model consist of a mean-field theory that includes the entropic and steric constributions of the polyelectrolyte and the electrostatic interactions between the polyelectrolyte and the capsid.  We initially model the genome as a flexible linear polyelectrolyte that interacts attractively with the positive charges residing on the binding domains and postpone the discussion of the impact of RNA secondary structure.

The free energy of a confined polyelectrolyte confined in a salt solution interacting with an external charge distribution can, within the ground-state approximation, be written as
\begin{multline} \label{free_energy}
  \beta F = \!\!\int\!\! {\mathrm{d}^3}{{\mathbf{r}}}\Big[
    \frac{1}{6} a^2 |{\nabla\Psi({\mathbf{r}})}|^2
    +\frac{1}{2}\upsilon \Psi^4 ({\mathbf{r}})\\
    -\frac{1}{8\pi\lambda_B} |{\nabla \beta e \Phi({\mathbf{r}})}|^2
    -2\mu\cosh\big[\beta e \Phi({\mathbf{r}})\big]\\
    + \beta \tau \Psi^2({\mathbf{r}}) \Phi({\mathbf{r}})
  + \beta \rho_0({\mathbf{r}}) \Phi({\mathbf{r}}) \Big],
\end{multline}
with $\beta$ the reciprocal temperature in units of energy, $a$ the statistical step or Kuhn length of the polymer, $v$ is effective excluded volume per monomer, $\lambda_B={e^2 \beta}/{4 \pi \epsilon}$ the Bjerrum length, $\epsilon$ the dielectric permittivity, $e$ the elementary charge, $\mu$ density of monovalent salt ions, and $\tau$ the linear charge density of chain.
The fields $\Psi(\mathbf{r})$ and $\Phi({\mathbf{r}})$ are the monomer density field and electrostatic potential of mean force respectively. The positive charge density $\rho_0({\mathbf{r}})$ is placed in an icosohedrally symetric distribution either on the capsid surface as shown in Figs.~1(a) and 1(b) or extending into the interior of the capsid along the N-terminal tails as in Fig.~1(c).
Extremizing the free energy with respect to the $\Psi({\mathbf r})$ and $\Phi({\mathbf r})$ fields subject to the constraint that the total number of monomers inside the capsid is constant \cite{Hone},
$N= \int {\mathrm{d}^3}{\mathbf r} \; \Psi^2 ({\mathbf{r}})$, 
results in two self-consistent non-linear field equations,
\begin{subequations} \label{eq:diff}
  \begin{align}
\frac{a^2}6\nabla^2\Psi & = -\mathcal{E}\Psi({\mathbf r})+\beta\tau \Phi({\mathbf r})\Psi({\mathbf r})+\upsilon\Psi^3\label{eq:diff_a}\\
\tfrac{\beta e^2}{4\pi \lambda_B} \nabla^2\Phi({\mathbf r}) & = +2\mu e\sinh \beta e\Phi({\mathbf r}) \! - \! \tau \Psi^2({\mathbf r}) \! - \! \rho({\mathbf r})\label{eq:diff_b},
\end{align}
\end{subequations}
with $\mathcal{E}$ the Lagrange multiplier enforcing the fixed number of monomers. Note $\rho({\mathbf r})$ here is the volume charge density that will be set to zero if there are no charges extended to the interior of capsid. The boundary conditions for the electrostatic potential inside and outside of the capsid that we model as a sphere of radius $R$ are,
\begin{subequations}\label{eq:BCEL}
\begin{align}
{\hat n\cdot}{\nabla \Phi_{in}}\mid_{r=R}-{\hat n\cdot}{\nabla \Phi_{out}}\mid_{r=R}&= {{4\pi\lambda_B}}{\sigma(\theta,\phi)}/\beta e^2\label{eq:BCEL_a}\\
\Phi_{in}(r)\mid_{r=R}&=\Phi_{out}(r)\mid_{r=R}\label{eq:BCEL_b}\\
\Phi_{out}(r)\mid_{r=\infty}&=0. 
\end{align}
\end{subequations}
with ${\sigma(\theta,\phi})$ the surface charge density. In case of a space charge distribution $\rho \neq 0$, but then we assume $\sigma = 0$. If the charges are localized to the surface, then $\sigma \neq 0$ but the volume charge density $\rho= 0$.  Thus if the charges from the capsid are lying completely on the capsid wall, the volume charge density $\rho({\mathbf r})=0$ in Eq.~\eqref{eq:diff_b}, and charge from the capsid is modeled as the surface charge $\sigma(\theta,\phi)$ in Eq.~\ref{eq:BCEL_a}. We use Dirichlet $\Psi(r)\mid_{r=R}=0$ boundary conditions for the chain density at the capsid wall but our findings are robust and we found the same results for Neumann boundary condition $\partial_r\Psi(r)\mid_{r=R}=0$.  Equation \ref{eq:diff_a} applies to a linear chain. 

To examine the combined effect of the secondary structure of RNA and non-uniform capsid charge distribution in this paper, we add to Eq.~\ref{free_energy} the following terms
\begin{align} \label{W_branched}
  -\frac{1}{\sqrt{a^3}}(f_e\Psi+\frac{a^3}{6}f_b\Psi^3),
\end{align}
which describe the statistics of an annealed branched polymer \cite{Lubensky,Nguyen-Bruinsma,Lee-Nguyen,Elleuch,adsorption2015,Gonca2014,Gonca2016} with $f_e$ and $f_b$ the fugacities of the end and branch points respectively (see Supporting Information (SI)).

To explicitly model N-terminal charged tails, we employ Icosahedral Symmetric Based Function(ISBF) for $T=1$ and $T=3$ structures with 60 and 180 positively charged regions, respectively. These functions are real-valued, complete, and orthogonal and can be written as a sum over spherical harmonics \cite{zheng2000},
\begin{align} 
   ISBF_{l,n}(\theta,\phi)=\sum_{m=-l}^{+l} b_{l,n,m}Y_{l,m}(\theta,\phi).
\end{align}
The ISBF functions are indexed by the integers $l$ and $n$, where $l(l+1)$ is the azimuthal separation constant, and $n \in \{0,1,...,N_l-1$\} indexes the different ISBFs and $N_l$ denotes the number of linearly independent ISBFs that can be constructed for a given $l$. The weights $b_{l,n,m}$ can be computed for each $l$ by comparing the expansion of icosahedrally symmetric set of delta functions in both spherical harmonics and ISBFs. 
\begin{figure}
\centering
\subfloat[]{\includegraphics[width=0.3\linewidth]{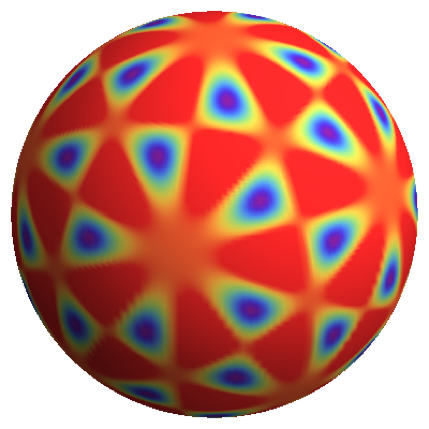}}\hspace{0.2cm}
\subfloat[]{\includegraphics[width=0.3\linewidth]{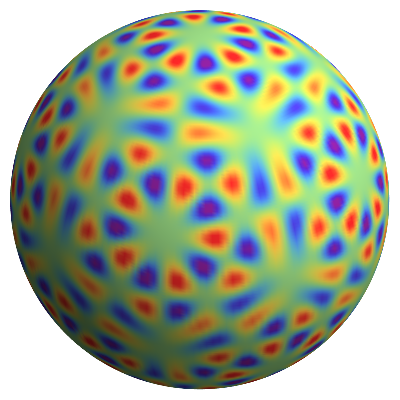}}\hspace{0.2cm}
\subfloat[]{\includegraphics[width=0.3\linewidth]{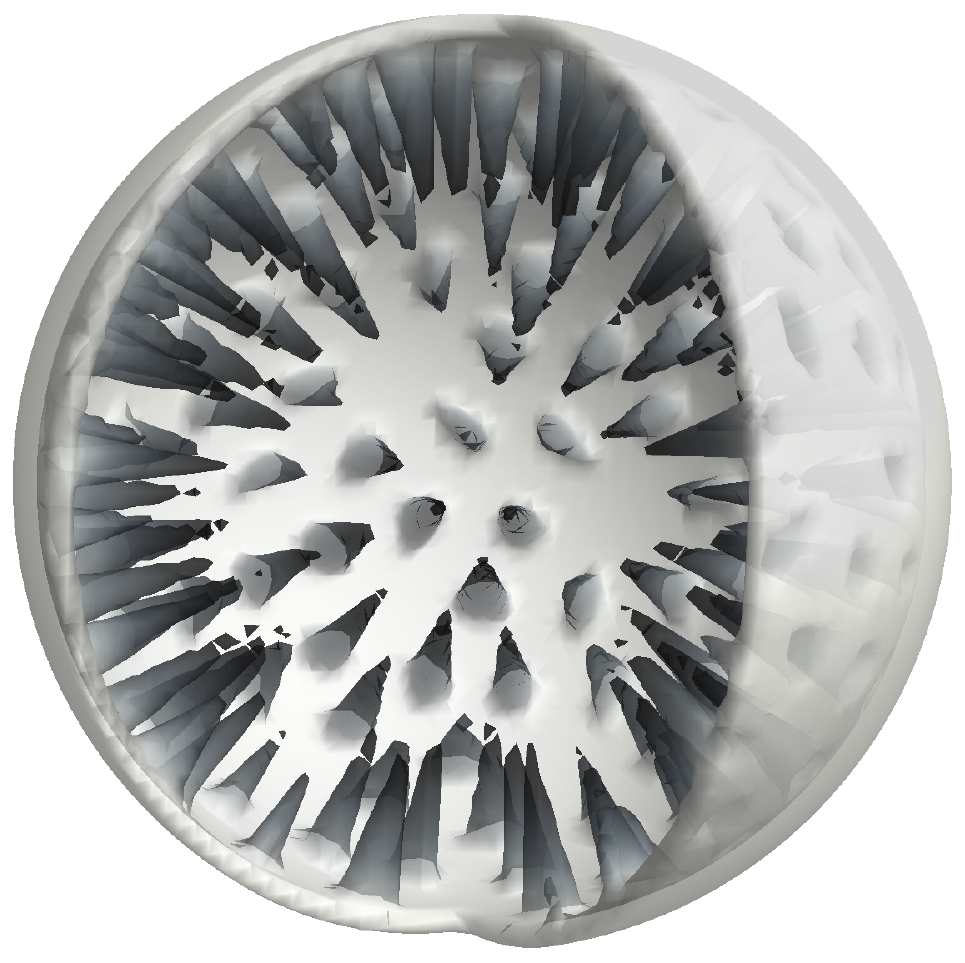}}
\caption{\footnotesize The charge distributions from the capsids for (a) a $T=1$ structure presented as $ISBF_{15,0}$ (b) a $T=3$ structure presented as $ISBF_{27,0}$ (c) and in the interior of the capsid inside a $T=3$ with 180 tails.}
\label{cap}
\end{figure}
The charge distributions for structures with $T=1$ and $T=3$ icosahedral symmetry are modeled by the ISBF with $(l=15, n=0)$ and $(l=27, n=0)$ respectively, shown in Figs.~1(a) and 1(b). The values of $b_{l,n,m}$ for T=1 and T=3 structures are given in SI. Assuming that there are no charges in the regions between N-terminal tails (see Fig. S1), we set charge density equal to zero if the magnitude of ISBFs is smaller than a certain cutoff value $C$. Thus, the distance between the charged regions depends on the cutoff, and, since we fix the total charge of the capsid, the charge density of the N-terminal domain changes as a function of the cutoff. We consider both the ``thin capsid model'' where the charges are smeared on the surface of the spherical capsid in 60 or 180 positions as shown in Figs.~1(a) and 1(b), and the ``thick capsid model'' where the charges extend inside the capsid as shown in Fig.~1(c). For the thick capsid model, we assumed that there are 60 $(T=1)$ or 180 $(T=3)$ ``bumpy'' charged regions extended inside the capsid. To this end, we shifted and truncated $ISBF_{15,0}$ and $ISBF_{27,0}$ such that the capsid surface protrudes in 60 or 180 positions presenting peptide tails (Fig.~1(c)) with charges uniformly distributed in the volume of protruded regions. 

We solved the coupled equations given in Eqs.~\ref{eq:diff} for $\Psi$ and $\Phi$ fields subject to the boundary conditions given in Eqs.~\ref{eq:BCEL} through the finite element method(FEM). The polymer density profiles $\Psi^2$ as a function of r, the distance from the center of the shell, are shown in Figs. S2 and S3 in one and three dimensions respectively. We find that the optimal genome length is increased for a non-uniform charge distribution compared to a uniform charge distribution and the free energy becomes deeper indicating the higher efficiency of genome encapsidation. Furthermore, we find that the optimal genome length increases if the cutoff $C$ is increased and that the distance between the charged regions is correspondingly increased. That is, as the charges on the capsid are distributed {\it more} non-uniformly, the optimal genome length increases. Nevertheless, for the thin capsid model, we have not been able to observe the phenomenon of overcharging with linear chains, {\it i.e.,} the number of charges on genome is always lower than those on the capsids for all the parameters that we tested. This is not the case for the thick capsid model as explained below.

\begin{figure} 
   \centering
   \includegraphics[width=0.45\textwidth]{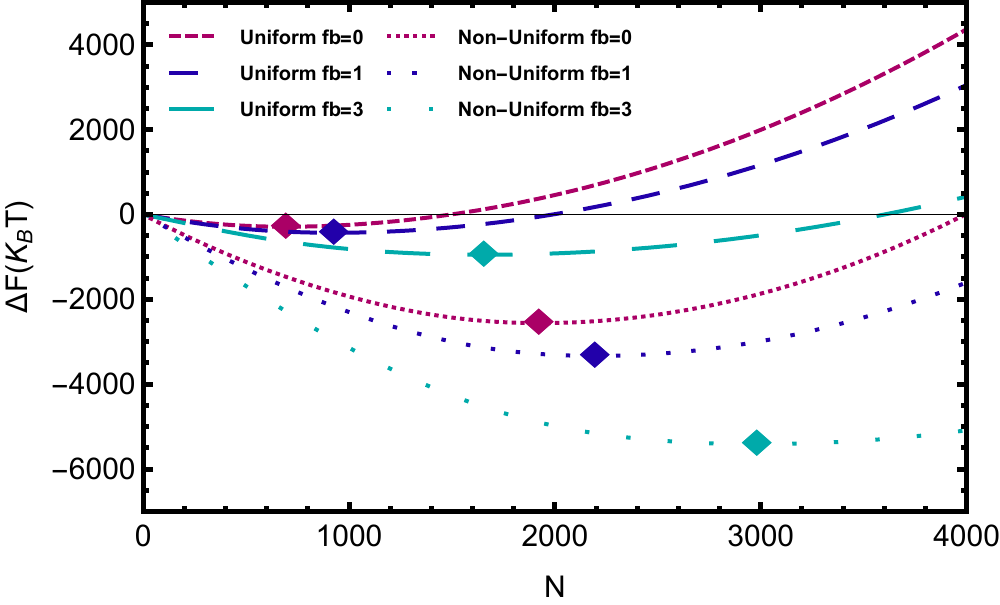}
   \caption{\footnotesize Encapsulation free energy for a linear and branched polyelectrolyte as a function of monomer number for a capsid with uniform (dashed lines) and non-uniform (dotted lines) charge density. For a linear chain the branching fugacity $f_b=0$ and increases to $f_b=1.0$ and $f_b=3.0$ as the chain becomes more branched. The diamonds indicate the minimum of free energy. Other parameters used correspond to a $T=3$ virus: total capsid charges on capsid $Q_c=1800$, $a=1.0$ $nm$, $v=0.01$ $nm^3$, $\mu=100$ $mM$,  $R=12$ $nm$, tail length $=4$ $nm$.}
   \label{FvsN}
\end{figure}

Figure \ref{FvsN} illustrates the encapsulation free energy as a function of genome length for a $T=3$ structure with the radius of capsid $R=12$ $nm$ at $\mu=100$ $mM$ salt concentrations for the thick capsid model (Fig.~1(c)).  The total number of charges are assumed to be $Q_c=1800$ for a $T=3$ structure, 10 charges on each N-terminal tail.  The dashed lines in Fig.~\ref{FvsN} correspond to a capsid with a uniform charge density and the dotted lines to a non-uniform charge density. The lines with the smallest distance between the dashed and dots correspond to a linear polymer. As illustrated in the figure, the minimum of the free energy moves towards longer chains when the charge distribution is non-uniform.

Figure \ref{FvsN} also shows the impact of RNA secondary structures on the optimal length of encapsidated genome.  The graphs in Fig. \ref{FvsN} corresponds to $f_b=0$ for a linear polymer and $f_b=1.0$ and $f_b=3.0$ for branched ones. The polymer becomes more branched as $f_b$ increases.  Note that in the figure the distance between dots or dashed lines increases as fugacity or the number of branch points increases. The figure reveals that as the degree of branching increases, the length of encapsidated genome increases for a capsid with a uniform charge density. This effect becomes stronger if we consider the capsid non-uniform charge distribution.  The diamonds in the figure indicate the optimal length of genome. The ratios of the optimal length or number of charges on RNA to the capsid total charge $Q_c=1800$ from left to right in the figure are 0.39, 0.52, 0.92, 1.07, 1.22, 1.66, which clearly shows a transition from undercharging towards overcharging. 
\begin{figure}
\centering
\subfloat[]{\includegraphics[width=1.\linewidth]{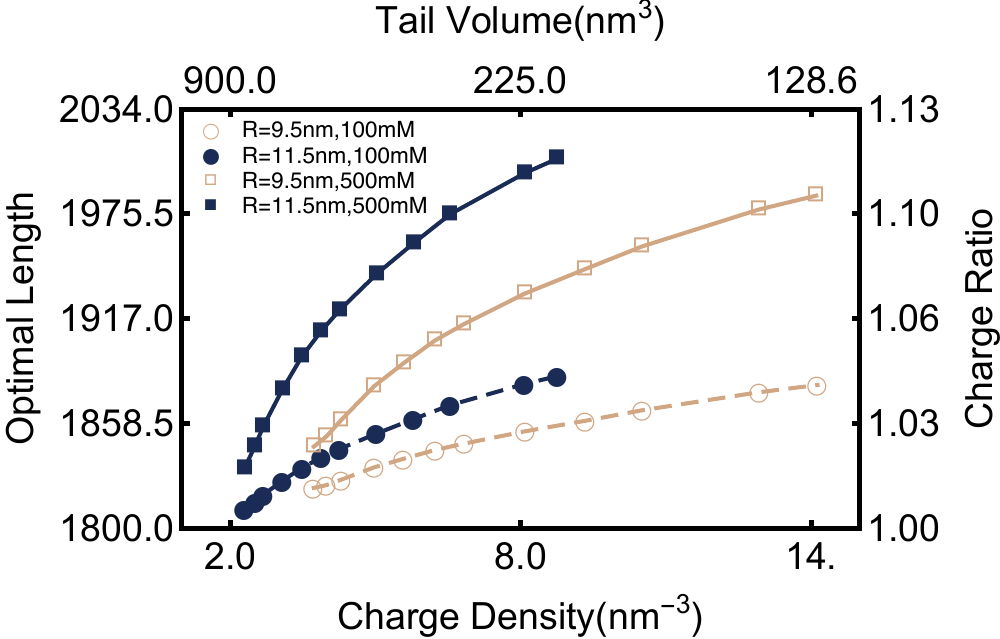}}\hfill
\subfloat[]{\includegraphics[width=1.\linewidth]{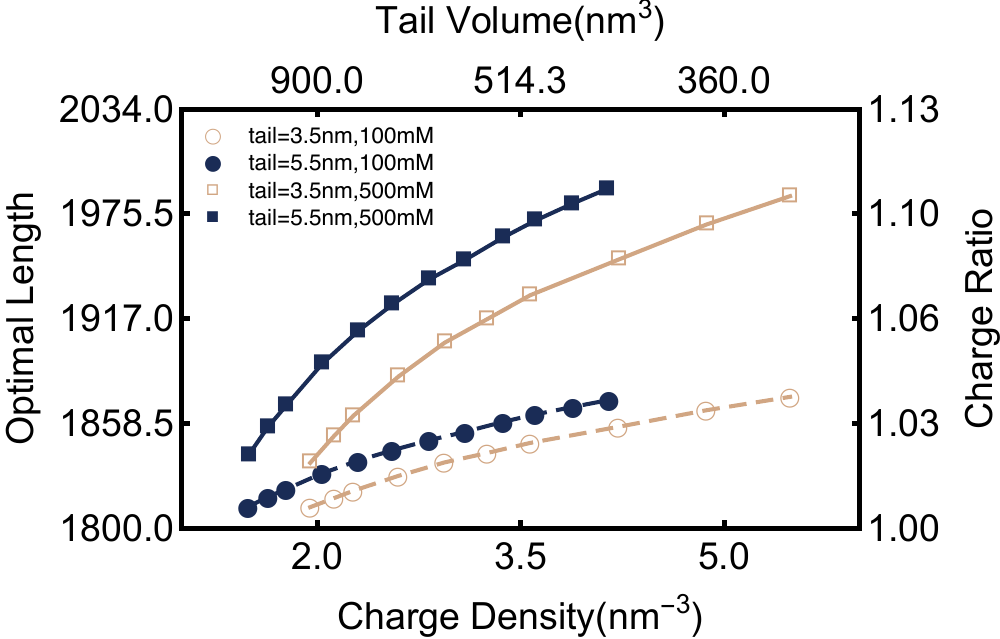}}
\caption{\footnotesize Optimal genome length or Charge Ratio vs. N-terminal charge density or volume occupied for a $T=3$. (a) Hollow symbols correspond to capsid radius of $R=9.5$ $nm$ and solid ones to $R=11.5$ $nm$ with the N-terminal tail length 3.5 $nm$; (b) Hollow symbols correspond to tail length of 3.5 $nm$ and solid ones to 5.5 $nm$ for $R=12.5$ $nm$. Other parameters are total charge $Qc=1800$, and salt concentration $\mu=100$ $mM$ (dashed) and $\mu=500$ $mM$ (solid).} \label{CRvsD}
\end{figure}

The reason for overcharging associated with the non-uniform charge distributions is two fold. The non-uniform charge distribution on the capsids obviously promotes non-uniform genome density. However, in order to have a more uniform polymer distribution with lower entropic cost, longer chains are preferably encapsidated to make the genome distribution more uniform in the regions between the N-terminal tails.  Figure 3(a) illustrates this effect for a $T=3$ structure with 180 tails as a plot of the optimal length of genome vs. capsid charge density. Note that since the total charge of capsid is fixed, as we increase the charge density, we lower the volume of the N-terminal regions, which is also shown in the axis on the top of the graph. The vertical axis on the right hand side of the figure shows the degree of overcharging.  The circles in the figure correspond to $\mu=100$ $mM$ salt concentration and squares to $\mu=500$ $mM$.  For the hollow symbols the radius of capsid is $R=9.5$ $nm$ but for the solid symbols $R=11.5$ $nm$.  As shown in the figure, if we increase the area between the N-terminals or the radius of capsid, the amount of overcharging increases at a given salt concentration.

However, the noted entropic effect cannot explain all the observations. At the physiological salt concentration ($\mu=100$ $mM$), the  genome only interacts with the capsid if it is sitting in vicinity of the capsid coat protein charges because of the rather short range of electrostatic interactions at high salt. The presence of N-terminals increases the region with which the genome interacts attractively through electrostatic interactions. Thus the higher the salt concentration, the more important becomes the role of N-terminals. The figure shows that the overcharging is more pronounced at $\mu=500$ $mM$.  Also the higher salt concentration, the lower is the electrostatic self-repulsion between genome monomers, which helps to the encapsidation of longer chains.

We also examined the impact of the length of N-terminal domains in the thick capsid model, which corresponds to how far the charged regions extend into the interior of the capsid. As illustrated in Fig.~3(b) for $T=3$ capsids, more genome is encapsidated for longer N-terminal tails, which is again due to larger interacting region for a fixed total numer of charges on the capsid. The effect becomes more pronounced for higher salt concentrations as illustrated in the figure. Finally, we investigated the effect of increasing the number of tails from 60 to 180 while keeping the radius of and the total number of charges on the capsid constant. As expected, if the number of tails increases, the optimal length of genome increases too.

In summary, we have studied the phenomena of overcharging observed in many viruses.  Previous mean-field theories as well as the experimental studies of CCMV capsid proteins with short linear polymers have indicated the resulting VLPs are undercharged\cite{siber2007,Siber2008,Ting,Gonca2014,Gonca2016,Siber2008,SiberZandi2010}. However, MD simulations revealed overcharging can happen even for linear polymers and the question is why\cite{elife}.  In this paper, we showed that the non-uniform charge distribution increases the amount of genome that can be assembled by capsid proteins due to entropic considerations. For a thin capsid model with the charges smeared flatly on the surface, longer chains are encapsidated but we have not been able to observe overcharging with linear polymers. This indicates that overcharging for linear systems is primarily due to the charged N-terminal regions that protrude into the interior of the capsid. The N-terminal regions increase the regions in which the genome can interact with the capsid proteins and thus resulting in the encapsidation of longer chains. This latter effect is stronger at higher salt concentrations. We find that the combined effect of RNA base-pairing which gives rise to the genome branching and non-uniform charge distribution can explain the pronounced charge inversion observed in viruses. 

The authors would like to thank Vladimir Lorman for useful
discussions. This work was supported by the National Science Foundation through Grant No. DMR-1310687 (R.Z.).


\bibliography{bibfile}

\begin{thebibliography}{40}%
\makeatletter
\providecommand \@ifxundefined [1]{%
 \@ifx{#1\undefined}
}%
\providecommand \@ifnum [1]{%
 \ifnum #1\expandafter \@firstoftwo
 \else \expandafter \@secondoftwo
 \fi
}%
\providecommand \@ifx [1]{%
 \ifx #1\expandafter \@firstoftwo
 \else \expandafter \@secondoftwo
 \fi
}%
\providecommand \natexlab [1]{#1}%
\providecommand \enquote  [1]{``#1''}%
\providecommand \bibnamefont  [1]{#1}%
\providecommand \bibfnamefont [1]{#1}%
\providecommand \citenamefont [1]{#1}%
\providecommand \href@noop [0]{\@secondoftwo}%
\providecommand \href [0]{\begingroup \@sanitize@url \@href}%
\providecommand \@href[1]{\@@startlink{#1}\@@href}%
\providecommand \@@href[1]{\endgroup#1\@@endlink}%
\providecommand \@sanitize@url [0]{\catcode `\\12\catcode `\$12\catcode
  `\&12\catcode `\#12\catcode `\^12\catcode `\_12\catcode `\%12\relax}%
\providecommand \@@startlink[1]{}%
\providecommand \@@endlink[0]{}%
\providecommand \url  [0]{\begingroup\@sanitize@url \@url }%
\providecommand \@url [1]{\endgroup\@href {#1}{\urlprefix }}%
\providecommand \urlprefix  [0]{URL }%
\providecommand \Eprint [0]{\href }%
\providecommand \doibase [0]{http://dx.doi.org/}%
\providecommand \selectlanguage [0]{\@gobble}%
\providecommand \bibinfo  [0]{\@secondoftwo}%
\providecommand \bibfield  [0]{\@secondoftwo}%
\providecommand \translation [1]{[#1]}%
\providecommand \BibitemOpen [0]{}%
\providecommand \bibitemStop [0]{}%
\providecommand \bibitemNoStop [0]{.\EOS\space}%
\providecommand \EOS [0]{\spacefactor3000\relax}%
\providecommand \BibitemShut  [1]{\csname bibitem#1\endcsname}%
\let\auto@bib@innerbib\@empty
\bibitem [{\citenamefont {Bancroft}(1970)}]{Bancroft}%
  \BibitemOpen
  \bibfield  {author} {\bibinfo {author} {\bibfnamefont {J.~B.}\ \bibnamefont
  {Bancroft}},\ }\href@noop {} {\bibfield  {journal} {\bibinfo  {journal} {Adv.
  Virus Res.}\ }\textbf {\bibinfo {volume} {16}},\ \bibinfo {pages} {99}
  (\bibinfo {year} {1970})}\BibitemShut {NoStop}%
\bibitem [{\citenamefont {Bruinsma}(2006)}]{bruinsma}%
  \BibitemOpen
  \bibfield  {author} {\bibinfo {author} {\bibfnamefont {R.~F.}\ \bibnamefont
  {Bruinsma}},\ }\href {\doibase 10.1140/epje/i2005-10071-1} {\bibfield
  {journal} {\bibinfo  {journal} {Euro. Phys. J. E}\ }\textbf {\bibinfo
  {volume} {19}},\ \bibinfo {pages} {303} (\bibinfo {year} {2006})}\BibitemShut
  {NoStop}%
\bibitem [{\citenamefont {Fejer}\ \emph {et~al.}(2010)\citenamefont {Fejer},
  \citenamefont {Chakrabarti},\ and\ \citenamefont {Wales}}]{Fejer:10}%
  \BibitemOpen
  \bibfield  {author} {\bibinfo {author} {\bibfnamefont {S.}~\bibnamefont
  {Fejer}}, \bibinfo {author} {\bibfnamefont {D.}~\bibnamefont {Chakrabarti}},
  \ and\ \bibinfo {author} {\bibfnamefont {D.}~\bibnamefont {Wales}},\
  }\href@noop {} {\bibfield  {journal} {\bibinfo  {journal} {Nano Lett}\
  }\textbf {\bibinfo {volume} {4}},\ \bibinfo {pages} {219} (\bibinfo {year}
  {2010})}\BibitemShut {NoStop}%
\bibitem [{\citenamefont {Rapaport}(2004)}]{Rapaport:04a}%
  \BibitemOpen
  \bibfield  {author} {\bibinfo {author} {\bibfnamefont {D.~C.}\ \bibnamefont
  {Rapaport}},\ }\href {\doibase 10.1103/PhysRevE.70.051905} {\bibfield
  {journal} {\bibinfo  {journal} {Phys Rev E}\ }\textbf {\bibinfo {volume}
  {70}},\ \bibinfo {pages} {051905} (\bibinfo {year} {2004})}\BibitemShut
  {NoStop}%
\bibitem [{\citenamefont {Wagner}\ and\ \citenamefont
  {Zandi}(2015)}]{Wagner2015956}%
  \BibitemOpen
  \bibfield  {author} {\bibinfo {author} {\bibfnamefont {J.}~\bibnamefont
  {Wagner}}\ and\ \bibinfo {author} {\bibfnamefont {R.}~\bibnamefont {Zandi}},\
  }\href {\doibase http://dx.doi.org/10.1016/j.bpj.2015.07.041} {\bibfield
  {journal} {\bibinfo  {journal} {Biophysical Journal}\ }\textbf {\bibinfo
  {volume} {109}},\ \bibinfo {pages} {956 } (\bibinfo {year}
  {2015})}\BibitemShut {NoStop}%
\bibitem [{\citenamefont {Chen}\ \emph {et~al.}(2007)\citenamefont {Chen},
  \citenamefont {Zhang},\ and\ \citenamefont {Glotzer}}]{Chen:2007b}%
  \BibitemOpen
  \bibfield  {author} {\bibinfo {author} {\bibfnamefont {T.}~\bibnamefont
  {Chen}}, \bibinfo {author} {\bibfnamefont {Z.}~\bibnamefont {Zhang}}, \ and\
  \bibinfo {author} {\bibfnamefont {S.~C.}\ \bibnamefont {Glotzer}},\
  }\href@noop {} {\bibfield  {journal} {\bibinfo  {journal} {PNAS}\ }\textbf
  {\bibinfo {volume} {104}},\ \bibinfo {pages} {717} (\bibinfo {year}
  {2007})}\BibitemShut {NoStop}%
\bibitem [{\citenamefont {Paquay}\ \emph {et~al.}(2016)\citenamefont {Paquay},
  \citenamefont {Kusumaatmaja}, \citenamefont {Wales}, \citenamefont {Zandi},\
  and\ \citenamefont {van~der Schoot}}]{Stefan}%
  \BibitemOpen
  \bibfield  {author} {\bibinfo {author} {\bibfnamefont {S.}~\bibnamefont
  {Paquay}}, \bibinfo {author} {\bibfnamefont {H.}~\bibnamefont
  {Kusumaatmaja}}, \bibinfo {author} {\bibfnamefont {D.~J.}\ \bibnamefont
  {Wales}}, \bibinfo {author} {\bibfnamefont {R.}~\bibnamefont {Zandi}}, \ and\
  \bibinfo {author} {\bibfnamefont {P.}~\bibnamefont {van~der Schoot}},\ }\href
  {\doibase 10.1039/C6SM00489J} {\bibfield  {journal} {\bibinfo  {journal}
  {Soft Matter}\ }\textbf {\bibinfo {volume} {12}},\ \bibinfo {pages} {5708}
  (\bibinfo {year} {2016})}\BibitemShut {NoStop}%
\bibitem [{\citenamefont {Sun}\ \emph {et~al.}(2007)\citenamefont {Sun},
  \citenamefont {DuFort}, \citenamefont {Daniel}, \citenamefont {Murali},
  \citenamefont {Chen}, \citenamefont {Gopinath}, \citenamefont {Stein},
  \citenamefont {De}, \citenamefont {Rotello}, \citenamefont {Holzenburg},
  \citenamefont {Kao},\ and\ \citenamefont {Dragnea}}]{Sun2007}%
  \BibitemOpen
  \bibfield  {author} {\bibinfo {author} {\bibfnamefont {J.}~\bibnamefont
  {Sun}}, \bibinfo {author} {\bibfnamefont {C.}~\bibnamefont {DuFort}},
  \bibinfo {author} {\bibfnamefont {M.-C.}\ \bibnamefont {Daniel}}, \bibinfo
  {author} {\bibfnamefont {A.}~\bibnamefont {Murali}}, \bibinfo {author}
  {\bibfnamefont {C.}~\bibnamefont {Chen}}, \bibinfo {author} {\bibfnamefont
  {K.}~\bibnamefont {Gopinath}}, \bibinfo {author} {\bibfnamefont
  {B.}~\bibnamefont {Stein}}, \bibinfo {author} {\bibfnamefont
  {M.}~\bibnamefont {De}}, \bibinfo {author} {\bibfnamefont {V.~M.}\
  \bibnamefont {Rotello}}, \bibinfo {author} {\bibfnamefont {A.}~\bibnamefont
  {Holzenburg}}, \bibinfo {author} {\bibfnamefont {C.~C.}\ \bibnamefont {Kao}},
  \ and\ \bibinfo {author} {\bibfnamefont {B.}~\bibnamefont {Dragnea}},\
  }\href@noop {} {\bibfield  {journal} {\bibinfo  {journal} {Proc. Nat. Acad.
  Sci. USA}\ }\textbf {\bibinfo {volume} {104}},\ \bibinfo {pages} {1354}
  (\bibinfo {year} {2007})}\BibitemShut {NoStop}%
\bibitem [{\citenamefont {Kusters}\ \emph {et~al.}(2015)\citenamefont
  {Kusters}, \citenamefont {Lin}, \citenamefont {Zandi}, \citenamefont
  {Tsvetkova}, \citenamefont {Dragnea},\ and\ \citenamefont {van~der
  Schoot}}]{Kusters2015}%
  \BibitemOpen
  \bibfield  {author} {\bibinfo {author} {\bibfnamefont {R.}~\bibnamefont
  {Kusters}}, \bibinfo {author} {\bibfnamefont {H.-K.}\ \bibnamefont {Lin}},
  \bibinfo {author} {\bibfnamefont {R.}~\bibnamefont {Zandi}}, \bibinfo
  {author} {\bibfnamefont {I.}~\bibnamefont {Tsvetkova}}, \bibinfo {author}
  {\bibfnamefont {B.}~\bibnamefont {Dragnea}}, \ and\ \bibinfo {author}
  {\bibfnamefont {P.}~\bibnamefont {van~der Schoot}},\ }\href {\doibase
  10.1021/jp5108125} {\bibfield  {journal} {\bibinfo  {journal} {J. Phys. Chem.
  B}\ }\textbf {\bibinfo {volume} {119}},\ \bibinfo {pages} {1869} (\bibinfo
  {year} {2015})}\BibitemShut {NoStop}%
\bibitem [{\citenamefont {Hagan}\ and\ \citenamefont
  {Zandi}(2016)}]{Zandi2016}%
  \BibitemOpen
  \bibfield  {author} {\bibinfo {author} {\bibfnamefont {M.~F.}\ \bibnamefont
  {Hagan}}\ and\ \bibinfo {author} {\bibfnamefont {R.}~\bibnamefont {Zandi}},\
  }\href {\doibase 10.1016/j.coviro.2016.02.012} {\bibfield  {journal}
  {\bibinfo  {journal} {Curr. Opin. Virol.}\ }\textbf {\bibinfo {volume}
  {18}},\ \bibinfo {pages} {36} (\bibinfo {year} {2016})}\BibitemShut {NoStop}%
\bibitem [{\citenamefont {Sikkema}\ \emph {et~al.}(2007)\citenamefont
  {Sikkema}, \citenamefont {Comellas-Aragones}, \citenamefont {Fokkink},
  \citenamefont {Verduin}, \citenamefont {Cornelissen},\ and\ \citenamefont
  {Nolte}}]{Cornelissen2007}%
  \BibitemOpen
  \bibfield  {author} {\bibinfo {author} {\bibfnamefont {F.~D.}\ \bibnamefont
  {Sikkema}}, \bibinfo {author} {\bibfnamefont {M.}~\bibnamefont
  {Comellas-Aragones}}, \bibinfo {author} {\bibfnamefont {R.~G.}\ \bibnamefont
  {Fokkink}}, \bibinfo {author} {\bibfnamefont {B.~J.~M.}\ \bibnamefont
  {Verduin}}, \bibinfo {author} {\bibfnamefont {J.}~\bibnamefont
  {Cornelissen}}, \ and\ \bibinfo {author} {\bibfnamefont {R.~J.~M.}\
  \bibnamefont {Nolte}},\ }\href {\doibase 10.1039/b613890j} {\bibfield
  {journal} {\bibinfo  {journal} {Org. Biomol. Chem.}\ }\textbf {\bibinfo
  {volume} {5}},\ \bibinfo {pages} {54} (\bibinfo {year} {2007})}\BibitemShut
  {NoStop}%
\bibitem [{\citenamefont {Ren}\ \emph {et~al.}(2006)\citenamefont {Ren},
  \citenamefont {Wong},\ and\ \citenamefont {Lim}}]{Ren2006}%
  \BibitemOpen
  \bibfield  {author} {\bibinfo {author} {\bibfnamefont {Y.~P.}\ \bibnamefont
  {Ren}}, \bibinfo {author} {\bibfnamefont {S.~M.}\ \bibnamefont {Wong}}, \
  and\ \bibinfo {author} {\bibfnamefont {L.~Y.}\ \bibnamefont {Lim}},\ }\href
  {\doibase 10.1099/vir.0.81944-0} {\bibfield  {journal} {\bibinfo  {journal}
  {J. Gen. Virol.}\ }\textbf {\bibinfo {volume} {87}},\ \bibinfo {pages} {2749}
  (\bibinfo {year} {2006})}\BibitemShut {NoStop}%
\bibitem [{\citenamefont {Ni}\ \emph {et~al.}(2012)\citenamefont {Ni},
  \citenamefont {Wang}, \citenamefont {Ma}, \citenamefont {Das}, \citenamefont
  {Sokol}, \citenamefont {Chiu}, \citenamefont {Dragnea}, \citenamefont
  {Hagan},\ and\ \citenamefont {Kao}}]{Bogdan}%
  \BibitemOpen
  \bibfield  {author} {\bibinfo {author} {\bibfnamefont {P.}~\bibnamefont
  {Ni}}, \bibinfo {author} {\bibfnamefont {Z.}~\bibnamefont {Wang}}, \bibinfo
  {author} {\bibfnamefont {X.}~\bibnamefont {Ma}}, \bibinfo {author}
  {\bibfnamefont {N.~C.}\ \bibnamefont {Das}}, \bibinfo {author} {\bibfnamefont
  {P.}~\bibnamefont {Sokol}}, \bibinfo {author} {\bibfnamefont
  {W.}~\bibnamefont {Chiu}}, \bibinfo {author} {\bibfnamefont {B.}~\bibnamefont
  {Dragnea}}, \bibinfo {author} {\bibfnamefont {M.}~\bibnamefont {Hagan}}, \
  and\ \bibinfo {author} {\bibfnamefont {C.~C.}\ \bibnamefont {Kao}},\ }\href
  {\doibase 10.1016/j.jmb.2012.03.023} {\bibfield  {journal} {\bibinfo
  {journal} {J. Mol. Biol.}\ }\textbf {\bibinfo {volume} {419}},\ \bibinfo
  {pages} {284} (\bibinfo {year} {2012})}\BibitemShut {NoStop}%
\bibitem [{\citenamefont {Losdorfer~Bozic}\ \emph {et~al.}(2013)\citenamefont
  {Losdorfer~Bozic}, \citenamefont {Siber},\ and\ \citenamefont
  {Podgornik}}]{Anze2}%
  \BibitemOpen
  \bibfield  {author} {\bibinfo {author} {\bibfnamefont {A.}~\bibnamefont
  {Losdorfer~Bozic}}, \bibinfo {author} {\bibfnamefont {A.}~\bibnamefont
  {Siber}}, \ and\ \bibinfo {author} {\bibfnamefont {R.}~\bibnamefont
  {Podgornik}},\ }\href@noop {} {\bibfield  {journal} {\bibinfo  {journal} {J.
  Biol. Phys.}\ }\textbf {\bibinfo {volume} {39}},\ \bibinfo {pages} {215}
  (\bibinfo {year} {2013})}\BibitemShut {NoStop}%
\bibitem [{\citenamefont {Daniel}\ \emph {et~al.}(2010)\citenamefont {Daniel},
  \citenamefont {Tsvetkova}, \citenamefont {Quinkert}, \citenamefont {Murali},
  \citenamefont {De}, \citenamefont {Rotello}, \citenamefont {Kao},\ and\
  \citenamefont {Dragnea}}]{DanielDragnea2010}%
  \BibitemOpen
  \bibfield  {author} {\bibinfo {author} {\bibfnamefont {M.~C.}\ \bibnamefont
  {Daniel}}, \bibinfo {author} {\bibfnamefont {I.~B.}\ \bibnamefont
  {Tsvetkova}}, \bibinfo {author} {\bibfnamefont {Z.~T.}\ \bibnamefont
  {Quinkert}}, \bibinfo {author} {\bibfnamefont {A.}~\bibnamefont {Murali}},
  \bibinfo {author} {\bibfnamefont {M.}~\bibnamefont {De}}, \bibinfo {author}
  {\bibfnamefont {V.~M.}\ \bibnamefont {Rotello}}, \bibinfo {author}
  {\bibfnamefont {C.~C.}\ \bibnamefont {Kao}}, \ and\ \bibinfo {author}
  {\bibfnamefont {B.}~\bibnamefont {Dragnea}},\ }\href {\doibase
  10.1021/nn1005073} {\bibfield  {journal} {\bibinfo  {journal} {ACS Nano}\
  }\textbf {\bibinfo {volume} {4}},\ \bibinfo {pages} {3853} (\bibinfo {year}
  {2010})}\BibitemShut {NoStop}%
\bibitem [{\citenamefont {Zlotnick}\ \emph {et~al.}(2000)\citenamefont
  {Zlotnick}, \citenamefont {Aldrich}, \citenamefont {Johnson}, \citenamefont
  {Ceres},\ and\ \citenamefont {Young}}]{Zlotnick}%
  \BibitemOpen
  \bibfield  {author} {\bibinfo {author} {\bibfnamefont {A.}~\bibnamefont
  {Zlotnick}}, \bibinfo {author} {\bibfnamefont {R.}~\bibnamefont {Aldrich}},
  \bibinfo {author} {\bibfnamefont {J.~M.}\ \bibnamefont {Johnson}}, \bibinfo
  {author} {\bibfnamefont {P.}~\bibnamefont {Ceres}}, \ and\ \bibinfo {author}
  {\bibfnamefont {M.~J.}\ \bibnamefont {Young}},\ }\href {\doibase
  10.1006/viro.2000.0619} {\bibfield  {journal} {\bibinfo  {journal}
  {Virology}\ }\textbf {\bibinfo {volume} {277}},\ \bibinfo {pages} {450}
  (\bibinfo {year} {2000})}\BibitemShut {NoStop}%
\bibitem [{\citenamefont {Lin}\ \emph {et~al.}(2012)\citenamefont {Lin},
  \citenamefont {van~der Schoot},\ and\ \citenamefont {Zandi}}]{Hsiang-Ku}%
  \BibitemOpen
  \bibfield  {author} {\bibinfo {author} {\bibfnamefont {H.-K.}\ \bibnamefont
  {Lin}}, \bibinfo {author} {\bibfnamefont {P.}~\bibnamefont {van~der Schoot}},
  \ and\ \bibinfo {author} {\bibfnamefont {R.}~\bibnamefont {Zandi}},\ }\href
  {\doibase 10.1088/1478-3975/9/6/066004} {\bibfield  {journal} {\bibinfo
  {journal} {Phys. Biol.}\ }\textbf {\bibinfo {volume} {9}},\ \bibinfo {pages}
  {066004} (\bibinfo {year} {2012})}\BibitemShut {NoStop}%
\bibitem [{\citenamefont {Sivanandam}\ \emph {et~al.}(2016)\citenamefont
  {Sivanandam}, \citenamefont {Mathews}, \citenamefont {Garmann}, \citenamefont
  {Erdemci-Tandogan}, \citenamefont {Zandi},\ and\ \citenamefont
  {Rao}}]{Venky2016}%
  \BibitemOpen
  \bibfield  {author} {\bibinfo {author} {\bibfnamefont {V.}~\bibnamefont
  {Sivanandam}}, \bibinfo {author} {\bibfnamefont {D.}~\bibnamefont {Mathews}},
  \bibinfo {author} {\bibfnamefont {R.}~\bibnamefont {Garmann}}, \bibinfo
  {author} {\bibfnamefont {G.}~\bibnamefont {Erdemci-Tandogan}}, \bibinfo
  {author} {\bibfnamefont {R.}~\bibnamefont {Zandi}}, \ and\ \bibinfo {author}
  {\bibfnamefont {A.~L.~N.}\ \bibnamefont {Rao}},\ }\href
  {http://dx.doi.org/10.1038/srep26328 http://10.1038/srep26328} {\bibfield
  {journal} {\bibinfo  {journal} {Scientific Reports}\ }\textbf {\bibinfo
  {volume} {6}},\ \bibinfo {pages} {26328} (\bibinfo {year}
  {2016})}\BibitemShut {NoStop}%
\bibitem [{\citenamefont {Comas-Garcia}\ \emph {et~al.}(2012)\citenamefont
  {Comas-Garcia}, \citenamefont {Cadena-Nava}, \citenamefont {Rao},
  \citenamefont {Knobler},\ and\ \citenamefont {Gelbart}}]{Comas}%
  \BibitemOpen
  \bibfield  {author} {\bibinfo {author} {\bibfnamefont {M.}~\bibnamefont
  {Comas-Garcia}}, \bibinfo {author} {\bibfnamefont {R.~D.}\ \bibnamefont
  {Cadena-Nava}}, \bibinfo {author} {\bibfnamefont {A.~L.~N.}\ \bibnamefont
  {Rao}}, \bibinfo {author} {\bibfnamefont {C.~M.}\ \bibnamefont {Knobler}}, \
  and\ \bibinfo {author} {\bibfnamefont {W.~M.}\ \bibnamefont {Gelbart}},\
  }\href {\doibase 10.1128/jvi.01695-12} {\bibfield  {journal} {\bibinfo
  {journal} {J. Virol.}\ }\textbf {\bibinfo {volume} {86}},\ \bibinfo {pages}
  {12271} (\bibinfo {year} {2012})}\BibitemShut {NoStop}%
\bibitem [{\citenamefont {Hu}\ \emph {et~al.}(2008)\citenamefont {Hu},
  \citenamefont {Zandi}, \citenamefont {Anavitarte}, \citenamefont {Knobler},\
  and\ \citenamefont {Gelbart}}]{Hu2008}%
  \BibitemOpen
  \bibfield  {author} {\bibinfo {author} {\bibfnamefont {Y.~F.}\ \bibnamefont
  {Hu}}, \bibinfo {author} {\bibfnamefont {R.}~\bibnamefont {Zandi}}, \bibinfo
  {author} {\bibfnamefont {A.}~\bibnamefont {Anavitarte}}, \bibinfo {author}
  {\bibfnamefont {C.~M.}\ \bibnamefont {Knobler}}, \ and\ \bibinfo {author}
  {\bibfnamefont {W.~M.}\ \bibnamefont {Gelbart}},\ }\href {\doibase
  10.1529/biophysj.107.117473} {\bibfield  {journal} {\bibinfo  {journal}
  {Biophys. J.}\ }\textbf {\bibinfo {volume} {94}},\ \bibinfo {pages} {1428}
  (\bibinfo {year} {2008})}\BibitemShut {NoStop}%
\bibitem [{\citenamefont {Cadena-Nava}\ \emph {et~al.}(2011)\citenamefont
  {Cadena-Nava}, \citenamefont {Hu}, \citenamefont {Garmann}, \citenamefont
  {Ng}, \citenamefont {Zelikin}, \citenamefont {Knobler},\ and\ \citenamefont
  {Gelbart}}]{Cadena2011}%
  \BibitemOpen
  \bibfield  {author} {\bibinfo {author} {\bibfnamefont {R.~D.}\ \bibnamefont
  {Cadena-Nava}}, \bibinfo {author} {\bibfnamefont {Y.~F.}\ \bibnamefont {Hu}},
  \bibinfo {author} {\bibfnamefont {R.~F.}\ \bibnamefont {Garmann}}, \bibinfo
  {author} {\bibfnamefont {B.}~\bibnamefont {Ng}}, \bibinfo {author}
  {\bibfnamefont {A.~N.}\ \bibnamefont {Zelikin}}, \bibinfo {author}
  {\bibfnamefont {C.~M.}\ \bibnamefont {Knobler}}, \ and\ \bibinfo {author}
  {\bibfnamefont {W.~M.}\ \bibnamefont {Gelbart}},\ }\href {\doibase
  10.1021/jp1094118} {\bibfield  {journal} {\bibinfo  {journal} {J. Phys. Chem.
  B}\ }\textbf {\bibinfo {volume} {115}},\ \bibinfo {pages} {2386} (\bibinfo
  {year} {2011})}\BibitemShut {NoStop}%
\bibitem [{\citenamefont {Perlmutter}\ \emph {et~al.}(2013)\citenamefont
  {Perlmutter}, \citenamefont {Qiao},\ and\ \citenamefont {Hagan}}]{elife}%
  \BibitemOpen
  \bibfield  {author} {\bibinfo {author} {\bibfnamefont {J.~D.}\ \bibnamefont
  {Perlmutter}}, \bibinfo {author} {\bibfnamefont {C.}~\bibnamefont {Qiao}}, \
  and\ \bibinfo {author} {\bibfnamefont {M.~F.}\ \bibnamefont {Hagan}},\ }\href
  {\doibase 10.7554/eLife.00632} {\bibfield  {journal} {\bibinfo  {journal}
  {eLife}\ }\textbf {\bibinfo {volume} {2}} (\bibinfo {year} {2013}),\
  10.7554/eLife.00632}\BibitemShut {NoStop}%
\bibitem [{\citenamefont {Singaram}\ \emph {et~al.}(2015)\citenamefont
  {Singaram}, \citenamefont {Garmann}, \citenamefont {Knobler}, \citenamefont
  {Gelbart},\ and\ \citenamefont {Ben-Shaul}}]{Ben-Shaul2015}%
  \BibitemOpen
  \bibfield  {author} {\bibinfo {author} {\bibfnamefont {S.~W.}\ \bibnamefont
  {Singaram}}, \bibinfo {author} {\bibfnamefont {R.~F.}\ \bibnamefont
  {Garmann}}, \bibinfo {author} {\bibfnamefont {C.~M.}\ \bibnamefont
  {Knobler}}, \bibinfo {author} {\bibfnamefont {W.~M.}\ \bibnamefont
  {Gelbart}}, \ and\ \bibinfo {author} {\bibfnamefont {A.}~\bibnamefont
  {Ben-Shaul}},\ }\href@noop {} {\bibfield  {journal} {\bibinfo  {journal}
  {Accounts of Chemical Research}\ }\textbf {\bibinfo {volume} {119}},\
  \bibinfo {pages} {13991} (\bibinfo {year} {2015})}\BibitemShut {NoStop}%
\bibitem [{\citenamefont {Erdemci-Tandogan}\ \emph {et~al.}(2014)\citenamefont
  {Erdemci-Tandogan}, \citenamefont {Wagner}, \citenamefont {van~der Schoot},
  \citenamefont {Podgornik},\ and\ \citenamefont {Zandi}}]{Gonca2014}%
  \BibitemOpen
  \bibfield  {author} {\bibinfo {author} {\bibfnamefont {G.}~\bibnamefont
  {Erdemci-Tandogan}}, \bibinfo {author} {\bibfnamefont {J.}~\bibnamefont
  {Wagner}}, \bibinfo {author} {\bibfnamefont {P.}~\bibnamefont {van~der
  Schoot}}, \bibinfo {author} {\bibfnamefont {R.}~\bibnamefont {Podgornik}}, \
  and\ \bibinfo {author} {\bibfnamefont {R.}~\bibnamefont {Zandi}},\ }\href
  {\doibase 10.1103/PhysRevE.89.032707} {\bibfield  {journal} {\bibinfo
  {journal} {Phys. Rev. E}\ }\textbf {\bibinfo {volume} {89}},\ \bibinfo
  {pages} {032707} (\bibinfo {year} {2014})}\BibitemShut {NoStop}%
\bibitem [{\citenamefont {Erdemci-Tandogan}\ \emph {et~al.}(2016)\citenamefont
  {Erdemci-Tandogan}, \citenamefont {Wagner}, \citenamefont {van~der Schoot},
  \citenamefont {Podgornik},\ and\ \citenamefont {Zandi}}]{Gonca2016}%
  \BibitemOpen
  \bibfield  {author} {\bibinfo {author} {\bibfnamefont {G.}~\bibnamefont
  {Erdemci-Tandogan}}, \bibinfo {author} {\bibfnamefont {J.}~\bibnamefont
  {Wagner}}, \bibinfo {author} {\bibfnamefont {P.}~\bibnamefont {van~der
  Schoot}}, \bibinfo {author} {\bibfnamefont {R.}~\bibnamefont {Podgornik}}, \
  and\ \bibinfo {author} {\bibfnamefont {R.}~\bibnamefont {Zandi}},\
  }\href@noop {} {\bibfield  {journal} {\bibinfo  {journal} {Phys. Rev. E}\
  }\textbf {\bibinfo {volume} {94}},\ \bibinfo {pages} {022408} (\bibinfo
  {year} {2016})}\BibitemShut {NoStop}%
\bibitem [{\citenamefont {Zandi}\ and\ \citenamefont {van~der
  Schoot}(2009)}]{Vanderschoot2009}%
  \BibitemOpen
  \bibfield  {author} {\bibinfo {author} {\bibfnamefont {R.}~\bibnamefont
  {Zandi}}\ and\ \bibinfo {author} {\bibfnamefont {P.}~\bibnamefont {van~der
  Schoot}},\ }\href {\doibase 10.1529/biophysj.108.137489} {\bibfield
  {journal} {\bibinfo  {journal} {Biophys. J.}\ }\textbf {\bibinfo {volume}
  {96}},\ \bibinfo {pages} {9} (\bibinfo {year} {2009})}\BibitemShut {NoStop}%
\bibitem [{\citenamefont {Bozic}\ \emph {et~al.}(2012)\citenamefont {Bozic},
  \citenamefont {Siber},\ and\ \citenamefont {Podgornik}}]{Anze}%
  \BibitemOpen
  \bibfield  {author} {\bibinfo {author} {\bibfnamefont {A.~L.}\ \bibnamefont
  {Bozic}}, \bibinfo {author} {\bibfnamefont {A.}~\bibnamefont {Siber}}, \ and\
  \bibinfo {author} {\bibfnamefont {R.}~\bibnamefont {Podgornik}},\ }\href
  {\doibase 10.1007/s10867-012-9278-4} {\bibfield  {journal} {\bibinfo
  {journal} {J. Biol. Phys.}\ }\textbf {\bibinfo {volume} {38}},\ \bibinfo
  {pages} {657} (\bibinfo {year} {2012})}\BibitemShut {NoStop}%
\bibitem [{\citenamefont {van~der Schoot}\ and\ \citenamefont
  {Bruinsma}(2005)}]{Vanderschoot}%
  \BibitemOpen
  \bibfield  {author} {\bibinfo {author} {\bibfnamefont {P.}~\bibnamefont
  {van~der Schoot}}\ and\ \bibinfo {author} {\bibfnamefont {R.}~\bibnamefont
  {Bruinsma}},\ }\href {\doibase 10.1103/PhysRevE.71.061928} {\bibfield
  {journal} {\bibinfo  {journal} {Phys. Rev. E}\ }\textbf {\bibinfo {volume}
  {71}},\ \bibinfo {pages} {061928} (\bibinfo {year} {2005})}\BibitemShut
  {NoStop}%
\bibitem [{\citenamefont {van~der Schoot}\ and\ \citenamefont
  {Zandi}(2013)}]{Paul13a}%
  \BibitemOpen
  \bibfield  {author} {\bibinfo {author} {\bibfnamefont {P.}~\bibnamefont
  {van~der Schoot}}\ and\ \bibinfo {author} {\bibfnamefont {R.}~\bibnamefont
  {Zandi}},\ }\href {\doibase 10.1007/s10867-013-9307-y} {\bibfield  {journal}
  {\bibinfo  {journal} {J. Biol. Phys.}\ }\textbf {\bibinfo {volume} {39}},\
  \bibinfo {pages} {289} (\bibinfo {year} {2013})}\BibitemShut {NoStop}%
\bibitem [{\citenamefont {Ji}\ and\ \citenamefont {Hone}(1988)}]{Hone}%
  \BibitemOpen
  \bibfield  {author} {\bibinfo {author} {\bibfnamefont {H.}~\bibnamefont
  {Ji}}\ and\ \bibinfo {author} {\bibfnamefont {D.}~\bibnamefont {Hone}},\
  }\href {\doibase 10.1021/ma00186a049} {\bibfield  {journal} {\bibinfo
  {journal} {Macromolecules}\ }\textbf {\bibinfo {volume} {21}},\ \bibinfo
  {pages} {2600} (\bibinfo {year} {1988})}\BibitemShut {NoStop}%
\bibitem [{\citenamefont {Lubensky}\ and\ \citenamefont
  {Isaacson}(1979)}]{Lubensky}%
  \BibitemOpen
  \bibfield  {author} {\bibinfo {author} {\bibfnamefont {T.~C.}\ \bibnamefont
  {Lubensky}}\ and\ \bibinfo {author} {\bibfnamefont {J.}~\bibnamefont
  {Isaacson}},\ }\href {\doibase 10.1103/PhysRevA.20.2130} {\bibfield
  {journal} {\bibinfo  {journal} {Phys. Rev. A}\ }\textbf {\bibinfo {volume}
  {20}},\ \bibinfo {pages} {2130} (\bibinfo {year} {1979})}\BibitemShut
  {NoStop}%
\bibitem [{\citenamefont {Nguyen}\ and\ \citenamefont
  {Bruinsma}(2006)}]{Nguyen-Bruinsma}%
  \BibitemOpen
  \bibfield  {author} {\bibinfo {author} {\bibfnamefont {T.~T.}\ \bibnamefont
  {Nguyen}}\ and\ \bibinfo {author} {\bibfnamefont {R.~F.}\ \bibnamefont
  {Bruinsma}},\ }\href {\doibase 10.1103/PhysRevLett.97.108102} {\bibfield
  {journal} {\bibinfo  {journal} {Phys. Rev. Lett.}\ }\textbf {\bibinfo
  {volume} {97}},\ \bibinfo {pages} {108102} (\bibinfo {year}
  {2006})}\BibitemShut {NoStop}%
\bibitem [{\citenamefont {Lee}\ and\ \citenamefont
  {Nguyen}(2008)}]{Lee-Nguyen}%
  \BibitemOpen
  \bibfield  {author} {\bibinfo {author} {\bibfnamefont {S.~I.}\ \bibnamefont
  {Lee}}\ and\ \bibinfo {author} {\bibfnamefont {T.~T.}\ \bibnamefont
  {Nguyen}},\ }\href {\doibase 10.1103/PhysRevLett.100.198102} {\bibfield
  {journal} {\bibinfo  {journal} {Phys. Rev. Lett.}\ }\textbf {\bibinfo
  {volume} {100}},\ \bibinfo {pages} {198102} (\bibinfo {year}
  {2008})}\BibitemShut {NoStop}%
\bibitem [{\citenamefont {Elleuch}\ \emph {et~al.}(1995)\citenamefont
  {Elleuch}, \citenamefont {Lequeux},\ and\ \citenamefont {Pfeuty}}]{Elleuch}%
  \BibitemOpen
  \bibfield  {author} {\bibinfo {author} {\bibfnamefont {K.}~\bibnamefont
  {Elleuch}}, \bibinfo {author} {\bibfnamefont {F.}~\bibnamefont {Lequeux}}, \
  and\ \bibinfo {author} {\bibfnamefont {P.}~\bibnamefont {Pfeuty}},\ }\href
  {\doibase 10.1051/jp1:1995140} {\bibfield  {journal} {\bibinfo  {journal} {J.
  Phys. I France}\ }\textbf {\bibinfo {volume} {5}},\ \bibinfo {pages} {465}
  (\bibinfo {year} {1995})}\BibitemShut {NoStop}%
\bibitem [{\citenamefont {Wagner}\ \emph {et~al.}(2015)\citenamefont {Wagner},
  \citenamefont {Erdemci-Tandogan},\ and\ \citenamefont
  {Zandi}}]{adsorption2015}%
  \BibitemOpen
  \bibfield  {author} {\bibinfo {author} {\bibfnamefont {J.}~\bibnamefont
  {Wagner}}, \bibinfo {author} {\bibfnamefont {G.}~\bibnamefont
  {Erdemci-Tandogan}}, \ and\ \bibinfo {author} {\bibfnamefont
  {R.}~\bibnamefont {Zandi}},\ }\href {\doibase 10.1088/0953-8984/27/49/495101}
  {\bibfield  {journal} {\bibinfo  {journal} {J. Phys.:Condens. Matter}\
  }\textbf {\bibinfo {volume} {27}},\ \bibinfo {pages} {495101} (\bibinfo
  {year} {2015})}\BibitemShut {NoStop}%
\bibitem [{\citenamefont {Zheng}\ and\ \citenamefont
  {Doerschuk}(2000)}]{zheng2000}%
  \BibitemOpen
  \bibfield  {author} {\bibinfo {author} {\bibfnamefont {Y.}~\bibnamefont
  {Zheng}}\ and\ \bibinfo {author} {\bibfnamefont {P.~C.}\ \bibnamefont
  {Doerschuk}},\ }\href {\doibase 10.1137/S0036141098341770} {\bibfield
  {journal} {\bibinfo  {journal} {SIAM Journal on Mathematical Analysis}\
  }\textbf {\bibinfo {volume} {32}},\ \bibinfo {pages} {538} (\bibinfo {year}
  {2000})},\ \Eprint
  {http://arxiv.org/abs/http://dx.doi.org/10.1137/S0036141098341770}
  {http://dx.doi.org/10.1137/S0036141098341770} \BibitemShut {NoStop}%
\bibitem [{\citenamefont {\v{S}iber}\ and\ \citenamefont
  {Podgornik}(2007)}]{siber2007}%
  \BibitemOpen
  \bibfield  {author} {\bibinfo {author} {\bibfnamefont {A.}~\bibnamefont
  {\v{S}iber}}\ and\ \bibinfo {author} {\bibfnamefont {R.}~\bibnamefont
  {Podgornik}},\ }\href {\doibase 10.1103/PhysRevE.76.061906} {\bibfield
  {journal} {\bibinfo  {journal} {Phys. Rev. E}\ }\textbf {\bibinfo {volume}
  {76}},\ \bibinfo {pages} {061906} (\bibinfo {year} {2007})},\ \Eprint
  {http://arxiv.org/abs/0709.0418} {arXiv:0709.0418} \BibitemShut {NoStop}%
\bibitem [{\citenamefont {Siber}\ and\ \citenamefont
  {Podgornik}(2008)}]{Siber2008}%
  \BibitemOpen
  \bibfield  {author} {\bibinfo {author} {\bibfnamefont {A.}~\bibnamefont
  {Siber}}\ and\ \bibinfo {author} {\bibfnamefont {R.}~\bibnamefont
  {Podgornik}},\ }\href {\doibase 10.1103/PhysRevE.78.051915} {\bibfield
  {journal} {\bibinfo  {journal} {Phys. Rev. E}\ }\textbf {\bibinfo {volume}
  {78}},\ \bibinfo {pages} {051915} (\bibinfo {year} {2008})}\BibitemShut
  {NoStop}%
\bibitem [{\citenamefont {Ting}\ \emph {et~al.}(2011)\citenamefont {Ting},
  \citenamefont {Wu},\ and\ \citenamefont {Wang}}]{Ting}%
  \BibitemOpen
  \bibfield  {author} {\bibinfo {author} {\bibfnamefont {C.~L.}\ \bibnamefont
  {Ting}}, \bibinfo {author} {\bibfnamefont {J.~Z.}\ \bibnamefont {Wu}}, \ and\
  \bibinfo {author} {\bibfnamefont {Z.~G.}\ \bibnamefont {Wang}},\ }\href
  {\doibase 10.1073/pnas.1109307108} {\bibfield  {journal} {\bibinfo  {journal}
  {PNAS}\ }\textbf {\bibinfo {volume} {108}},\ \bibinfo {pages} {16986}
  (\bibinfo {year} {2011})}\BibitemShut {NoStop}%
\bibitem [{\citenamefont {Siber}\ \emph {et~al.}(2010)\citenamefont {Siber},
  \citenamefont {Zandi},\ and\ \citenamefont {Podgornik}}]{SiberZandi2010}%
  \BibitemOpen
  \bibfield  {author} {\bibinfo {author} {\bibfnamefont {A.}~\bibnamefont
  {Siber}}, \bibinfo {author} {\bibfnamefont {R.}~\bibnamefont {Zandi}}, \ and\
  \bibinfo {author} {\bibfnamefont {R.}~\bibnamefont {Podgornik}},\ }\href
  {\doibase 10.1103/PhysRevE.81.051919} {\bibfield  {journal} {\bibinfo
  {journal} {Phys. Rev. E}\ }\textbf {\bibinfo {volume} {81}},\ \bibinfo
  {pages} {051919} (\bibinfo {year} {2010})}\BibitemShut {NoStop}%
\end{thebibliography}%


\begin{thebibliography}{7}%
\makeatletter
\providecommand \@ifxundefined [1]{%
 \@ifx{#1\undefined}
}%
\providecommand \@ifnum [1]{%
 \ifnum #1\expandafter \@firstoftwo
 \else \expandafter \@secondoftwo
 \fi
}%
\providecommand \@ifx [1]{%
 \ifx #1\expandafter \@firstoftwo
 \else \expandafter \@secondoftwo
 \fi
}%
\providecommand \natexlab [1]{#1}%
\providecommand \enquote  [1]{``#1''}%
\providecommand \bibnamefont  [1]{#1}%
\providecommand \bibfnamefont [1]{#1}%
\providecommand \citenamefont [1]{#1}%
\providecommand \href@noop [0]{\@secondoftwo}%
\providecommand \href [0]{\begingroup \@sanitize@url \@href}%
\providecommand \@href[1]{\@@startlink{#1}\@@href}%
\providecommand \@@href[1]{\endgroup#1\@@endlink}%
\providecommand \@sanitize@url [0]{\catcode `\\12\catcode `\$12\catcode
  `\&12\catcode `\#12\catcode `\^12\catcode `\_12\catcode `\%12\relax}%
\providecommand \@@startlink[1]{}%
\providecommand \@@endlink[0]{}%
\providecommand \url  [0]{\begingroup\@sanitize@url \@url }%
\providecommand \@url [1]{\endgroup\@href {#1}{\urlprefix }}%
\providecommand \urlprefix  [0]{URL }%
\providecommand \Eprint [0]{\href }%
\providecommand \doibase [0]{http://dx.doi.org/}%
\providecommand \selectlanguage [0]{\@gobble}%
\providecommand \bibinfo  [0]{\@secondoftwo}%
\providecommand \bibfield  [0]{\@secondoftwo}%
\providecommand \translation [1]{[#1]}%
\providecommand \BibitemOpen [0]{}%
\providecommand \bibitemStop [0]{}%
\providecommand \bibitemNoStop [0]{.\EOS\space}%
\providecommand \EOS [0]{\spacefactor3000\relax}%
\providecommand \BibitemShut  [1]{\csname bibitem#1\endcsname}%
\let\auto@bib@innerbib\@empty
\bibitem [{\citenamefont {Lubensky}\ and\ \citenamefont
  {Isaacson}(1979)}]{Lubensky}%
  \BibitemOpen
  \bibfield  {author} {\bibinfo {author} {\bibfnamefont {T.~C.}\ \bibnamefont
  {Lubensky}}\ and\ \bibinfo {author} {\bibfnamefont {J.}~\bibnamefont
  {Isaacson}},\ }\href {\doibase 10.1103/PhysRevA.20.2130} {\bibfield
  {journal} {\bibinfo  {journal} {Phys. Rev. A}\ }\textbf {\bibinfo {volume}
  {20}},\ \bibinfo {pages} {2130} (\bibinfo {year} {1979})}\BibitemShut
  {NoStop}%
\bibitem [{\citenamefont {Nguyen}\ and\ \citenamefont
  {Bruinsma}(2006)}]{Nguyen-Bruinsma}%
  \BibitemOpen
  \bibfield  {author} {\bibinfo {author} {\bibfnamefont {T.~T.}\ \bibnamefont
  {Nguyen}}\ and\ \bibinfo {author} {\bibfnamefont {R.~F.}\ \bibnamefont
  {Bruinsma}},\ }\href {\doibase 10.1103/PhysRevLett.97.108102} {\bibfield
  {journal} {\bibinfo  {journal} {Phys. Rev. Lett.}\ }\textbf {\bibinfo
  {volume} {97}},\ \bibinfo {pages} {108102} (\bibinfo {year}
  {2006})}\BibitemShut {NoStop}%
\bibitem [{\citenamefont {Lee}\ and\ \citenamefont
  {Nguyen}(2008)}]{Lee-Nguyen}%
  \BibitemOpen
  \bibfield  {author} {\bibinfo {author} {\bibfnamefont {S.~I.}\ \bibnamefont
  {Lee}}\ and\ \bibinfo {author} {\bibfnamefont {T.~T.}\ \bibnamefont
  {Nguyen}},\ }\href {\doibase 10.1103/PhysRevLett.100.198102} {\bibfield
  {journal} {\bibinfo  {journal} {Phys. Rev. Lett.}\ }\textbf {\bibinfo
  {volume} {100}},\ \bibinfo {pages} {198102} (\bibinfo {year}
  {2008})}\BibitemShut {NoStop}%
\bibitem [{\citenamefont {Elleuch}\ \emph {et~al.}(1995)\citenamefont
  {Elleuch}, \citenamefont {Lequeux},\ and\ \citenamefont {Pfeuty}}]{Elleuch}%
  \BibitemOpen
  \bibfield  {author} {\bibinfo {author} {\bibfnamefont {K.}~\bibnamefont
  {Elleuch}}, \bibinfo {author} {\bibfnamefont {F.}~\bibnamefont {Lequeux}}, \
  and\ \bibinfo {author} {\bibfnamefont {P.}~\bibnamefont {Pfeuty}},\ }\href
  {\doibase 10.1051/jp1:1995140} {\bibfield  {journal} {\bibinfo  {journal} {J.
  Phys. I France}\ }\textbf {\bibinfo {volume} {5}},\ \bibinfo {pages} {465}
  (\bibinfo {year} {1995})}\BibitemShut {NoStop}%
\bibitem [{\citenamefont {Erdemci-Tandogan}\ \emph {et~al.}(2014)\citenamefont
  {Erdemci-Tandogan}, \citenamefont {Wagner}, \citenamefont {van~der Schoot},
  \citenamefont {Podgornik},\ and\ \citenamefont {Zandi}}]{Gonca2014}%
  \BibitemOpen
  \bibfield  {author} {\bibinfo {author} {\bibfnamefont {G.}~\bibnamefont
  {Erdemci-Tandogan}}, \bibinfo {author} {\bibfnamefont {J.}~\bibnamefont
  {Wagner}}, \bibinfo {author} {\bibfnamefont {P.}~\bibnamefont {van~der
  Schoot}}, \bibinfo {author} {\bibfnamefont {R.}~\bibnamefont {Podgornik}}, \
  and\ \bibinfo {author} {\bibfnamefont {R.}~\bibnamefont {Zandi}},\ }\href
  {\doibase 10.1103/PhysRevE.89.032707} {\bibfield  {journal} {\bibinfo
  {journal} {Phys. Rev. E}\ }\textbf {\bibinfo {volume} {89}},\ \bibinfo
  {pages} {032707} (\bibinfo {year} {2014})}\BibitemShut {NoStop}%
\bibitem [{\citenamefont {Erdemci-Tandogan}\ \emph {et~al.}(2016)\citenamefont
  {Erdemci-Tandogan}, \citenamefont {Wagner}, \citenamefont {van~der Schoot},
  \citenamefont {Podgornik},\ and\ \citenamefont {Zandi}}]{Gonca2016}%
  \BibitemOpen
  \bibfield  {author} {\bibinfo {author} {\bibfnamefont {G.}~\bibnamefont
  {Erdemci-Tandogan}}, \bibinfo {author} {\bibfnamefont {J.}~\bibnamefont
  {Wagner}}, \bibinfo {author} {\bibfnamefont {P.}~\bibnamefont {van~der
  Schoot}}, \bibinfo {author} {\bibfnamefont {R.}~\bibnamefont {Podgornik}}, \
  and\ \bibinfo {author} {\bibfnamefont {R.}~\bibnamefont {Zandi}},\
  }\href@noop {} {\bibfield  {journal} {\bibinfo  {journal} {Phys. Rev. E}\
  }\textbf {\bibinfo {volume} {94}},\ \bibinfo {pages} {022408} (\bibinfo
  {year} {2016})}\BibitemShut {NoStop}%
\bibitem [{\citenamefont {Zheng}\ and\ \citenamefont
  {Doerschuk}(2000)}]{zheng2000}%
  \BibitemOpen
  \bibfield  {author} {\bibinfo {author} {\bibfnamefont {Y.}~\bibnamefont
  {Zheng}}\ and\ \bibinfo {author} {\bibfnamefont {P.~C.}\ \bibnamefont
  {Doerschuk}},\ }\href {\doibase 10.1137/S0036141098341770} {\bibfield
  {journal} {\bibinfo  {journal} {SIAM Journal on Mathematical Analysis}\
  }\textbf {\bibinfo {volume} {32}},\ \bibinfo {pages} {538} (\bibinfo {year}
  {2000})},\ \Eprint
  {http://arxiv.org/abs/http://dx.doi.org/10.1137/S0036141098341770}
  {http://dx.doi.org/10.1137/S0036141098341770} \BibitemShut {NoStop}%
\end{thebibliography}%


%
\end{document}


\preprint{APS/123-QED}
\title{The impact of non-uniform capsid charge density on virus assembly\\Supplementary Information}

\author{Siyu Li}
\author{Gonca Erdemci-Tandogan}
\author{Jef Wagner}
\affiliation{Department of Physics and Astronomy,
University of California, Riverside, California 92521, USA}
\author{Paul van der Schoot}
\affiliation{Group Theory of Polymers and Soft Matter, Eindhoven University of Technology, P.O. Box 513, 5600 MB Eindhoven,
The Netherlands}
 \affiliation{ Institute for Theoretical Physics,
   Utrecht University,
   Leuvenlaan 4, 3584 CE Utrecht, The Netherlands}
\author{Roya Zandi}
\affiliation{Department of Physics and Astronomy,
   University of California, Riverside, California 92521, USA}





\date{\today}

\maketitle

\section{\label{figs}MANUSCRIPT SUPPORTING FIGURES}

\begin{figure}[h]
\centering
\includegraphics[width=0.5\linewidth]{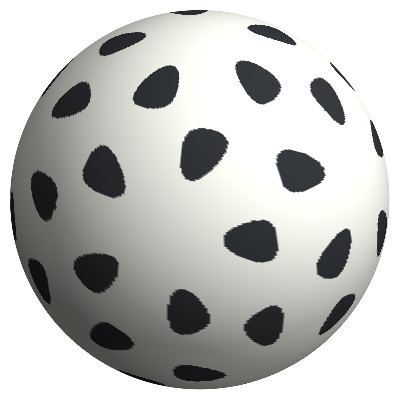}
\caption{\footnotesize A T=1 structure.  The black spots show the regions with a uniform surface charge density. The charges are smeared on the surface representing the thin capsid model.}
\label{CRvsD}
\end{figure}
\begin{figure}[h]
\centering
\includegraphics[width=0.85\linewidth]{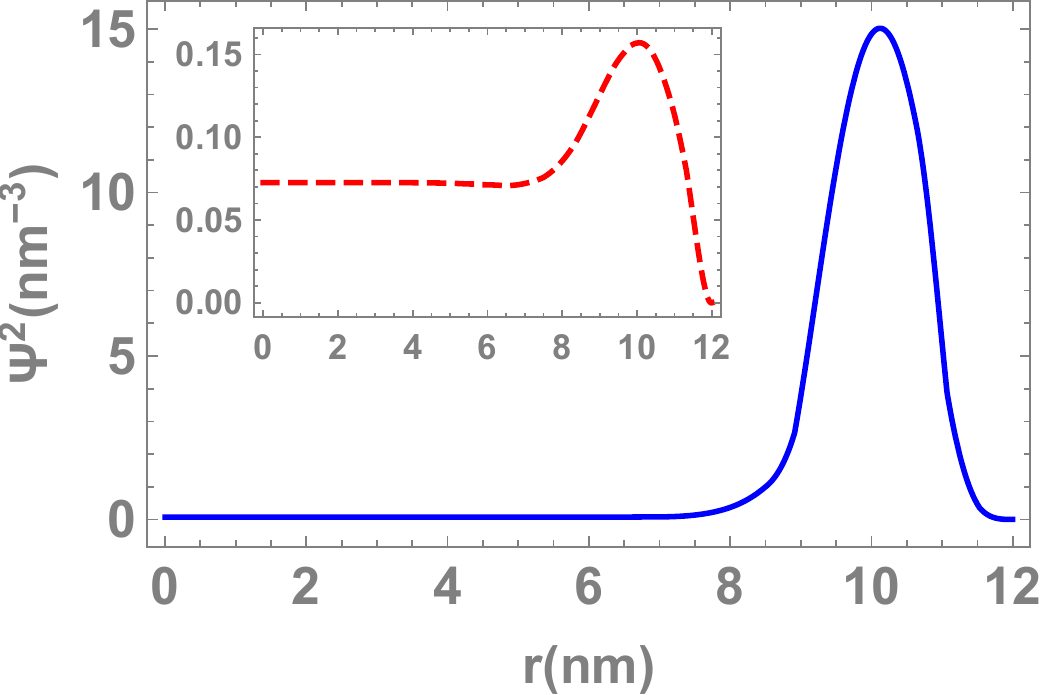}
\caption{\footnotesize Density profile vs. the capsid radius for a T=3 structure with non-uniform charge distribution. The figure shows the profiles along two different directions. The solid line corresponds to the direction in which the N-terminal tail is located and the dashed line to the direction without N-terminal tail (inset graph). In the absence of surface charge density and N-terminal tail (dashed curve), the density is still maximum close to the wall. The polymer is branched with $f_b=3$, total monomer number=2411, salt concentration $\mu=100$mM, $R=12$nm and $Qc=1800$. }
\label{branch profile}
\end{figure}
\begin{figure}[h]
\centering
\includegraphics[width=0.5\linewidth]{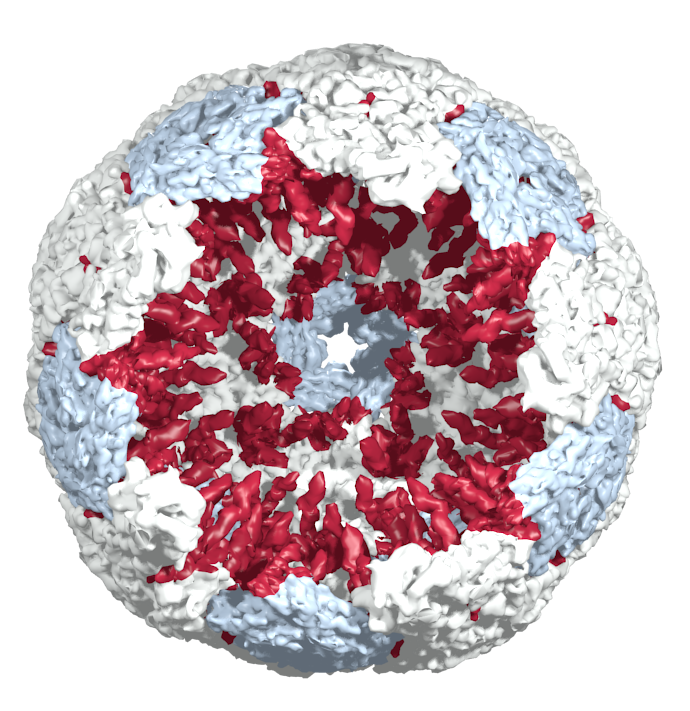}
\caption{\footnotesize 3D view of genome density profile in a T=3 capsid. The protruded regions represent RNA (red). The density of RNA between N-terminals is very small and not shown in the figure.}
\label{CRvsD}
\end{figure}

\section{\label{branch}Branched Polymer}
In a previous work, we showed that as the secondary structure of the genome becomes more branched more genome can be encapsidated giving rise to the phenomena of overcharging. To elucidate the impact of branching we modeled RNA as an annealed branched polymer by adding the following terms to the free energy(Eq.~1),
\begin{align} \label{W_branched}
   -\frac{1}{\sqrt{a^3}}(f_e\Psi+\frac{a^3}{6}f_b\Psi^3).
\end{align}
The field equations (Eq.~2) become
\begin{subequations} \label{eq:diff}
\begin{align}\notag
\frac{a^2}6\nabla^2\Psi&=-\mathcal{E}\Psi({\mathbf r})+ \beta\tau \Phi({\mathbf r})\Psi({\mathbf r})+\upsilon\Psi^3\\
&\hspace{2cm} -\frac{f_e}{2\sqrt{a^3}}-\frac{\sqrt{a^3}}{4}f_b\Psi^2,\label{eq:diff_a}\\
\tfrac{\beta e^2}{4\pi \lambda_B} \nabla^2\Phi({\mathbf r})&=+2\mu e\sinh \beta e\Phi({\mathbf r}) \! - \! \tau \Psi^2({\mathbf r}) \! - \! \rho({\mathbf r})\label{eq:diff_b},
\end{align}
\end{subequations}
which describe the statistics of an annealed branched polymer \cite{Lubensky,Nguyen-Bruinsma,Lee-Nguyen,Elleuch,Gonca2014,Gonca2016} with $f_e$ and $f_b$ the fugacities of the end and branch points, respectively.  In this formalism, the stem-loops or hair-pins in RNA are considered as end points. The number of end and branch points $N_e$ and $N_b$ of the polymer depend on the fugacities $N_e =- \beta f_e \frac{\partial{F}}{\partial{f_e}}$ and $N_b =- \beta f_b \frac{\partial{F}}{\partial{f_b}}$. We consider only the case of a single encapsulated polymer with no closed loops, and thus we have the following constraint:  $N_e = N_b+2$. The fugacity of branch points $f_b$ determines the degree of branching.

\section{\label{ISBF} Icosahedral Symmetric Based Function(ISBF)}
The coefficients, $b_{l,n,m}$ given in Eq.~5 become nonzero only when $m$ is a multiple of five, corresponding to five-fold symmetry of icosahedral group.  As a function of the associated Legendre function $P_m^l(x)$, ISBFs \cite{zheng2000} can easily be written as
\begin{eqnarray}
&&ISBF_{l,n}(\theta,\phi)=\notag\\
&&\left\{
\begin{array}{ll}\notag
\sum_{m=0}^{+l} \frac{2}{1+\delta_{m,0}}b_{l,n,m}N_{l,m}P_m^l(\cos \theta)\cos(m\phi),&l:even \\ 
\sum_{m=1}^{+l} 2b_{l,n,m}N_{l,m} P_m^l (\cos\theta)\sin(m\phi), &l:odd\\ 
\end{array}
\right.\\
\end{eqnarray}
\\
\begingroup\makeatletter\def\f@size{8}\check@mathfonts
with $N_{l,m}=\sqrt{\frac{2l+1}{4\pi}\frac{(l-m)!}{(l+m)!}}$.\\
The table S1 shows $b_{l,n,m}$ for $T=1$ and $T=3$ structures.

\begin{table}[H]
\caption{\label{tab:table1}The coefficients of $b_{l,n,m}$ for $ISBF_{15,0}$ ($T=1$) and $ISBF_{27,0}$ ($T=3$) structures, see Eq.~5 in the manuscript.}
\begin{ruledtabular}
\begin{tabular}{cccccc}
$l$& $b_{l,0,5}$ & $b_{l,0,10}$ & $b_{l,0,15}$ & $b_{l,0,20}$ & $b_{l,0,25}$\\
\colrule
15 & 0.51653 & 0.39131 & -0.28298 &  & \\
27 & 0.44330 & -0.23513 & -0.02788 & 0.41768 & -0.27011\\
\end{tabular}
\end{ruledtabular}
\end{table}
\newpage
\bibliography{bibfile}